\documentclass[aps,prl,reprint,groupedaddress,nofootinbib]{revtex4-2}
\usepackage{graphicx}

\begin{document}

%\preprint{}

\title{Photonic-computing error correction through optical en-/decoder calibrations}
\let\thefootnote\relax\footnotetext{* adamc@dtu.dk}
\let\thefootnote\relax\footnotetext{$^\dagger$ bala@dtu.dk}

\author{Adam Carstensen*}
%\email[]{adamc@dtu.dk}
\affiliation{Department of Electrical and Photonics Engineering, Technical University of Denmark, 2800 Kgs.\ Lyngby,
Denmark}
%\thanks{}
%

\author{Babak Vosoughi Lahijani$^\dagger$}
%\email[]{bala@dtu.dk}
\affiliation{Department of Electrical and Photonics Engineering, Technical University of Denmark, 2800 Kgs.\ Lyngby,
Denmark}


\date{\today}

\begin{abstract}
\noindent
Photonic processors have emerged as an attractive platform for fast and energy-efficient matrix-vector multiplication. However, they are susceptible to error due to their analog nature. Here, we present an error-correction technique that implements a correction offset to the optical en-/decoders of photonic processors. Our proposed method is general-purpose, does not require introducing any additional components to the photonic network, and can address errors stemming from unbalanced losses, 50/50 beamsplitter deviations, digital-to-analog conversion inaccuracies, and any unknown sources. In particular, we show that our method is highly effective in mitigating unbalanced-loss errors, a problem that has not previously been addressed by any error-correction technique. Using this approach, we achieve over 90\% error reduction in large triangular meshes, overcoming a key obstacle to highly accurate photonic processors for information processing.
\end{abstract}

%\keywords{}

\maketitle

% References should be done using the~\cite, \ref, and \label commands
\section{Introduction}
\noindent
Digital electronic processors have reached a plateau in computing efficiency, with only marginal improvements over the past decade~\cite{shalf2020future,rupp_microprocessor_trends}. At the same time, the demand for computation has increased dramatically, driven in large part by artificial intelligence (AI) workloads~\cite{strubell2020energy}. A large fraction of these workloads, such as neural network training and inference, are dominated by linear algebra operations, in particular matrix–vector multiplications (MVMs)~\cite{krizhevsky2012imagenet,brown2020language,silver2017mastering}. Beyond AI, linear operation is central to computing and forms the basis for a wide range of applications such as signal processing, communications, scientific simulations, and optimization. Therefore, fast and efficient MVM is a key objective for next-generation information processing systems.
Analog photonic computing offers a promising path towards this goal, as photonic devices can perform linear operations inherently and in parallel~\cite{mcmahon2023physics}. In particular, photonic matrix–vector multiplication processors (PMVMPs) can achieve computing speeds and energy efficiencies beyond what is possible with digital electronics~\cite{zhou2022photonic,harris2018linear,fu2024optical,ahmed2025universal,hua2025integrated}.
PMVMPs have been demonstrated using various approaches such as cross-bar architectures~\cite{moralis2024perfect}, spatial light modulators~\cite{wang2022optical,lin2018all}, unitary optical processors~\cite{tang2021ten,tang2017integrated}, wavelength multiplexing~\cite{xu202111}, and integrated Mach-Zehnder interferometers (MZI)~\cite{zhang2021optical,carolan2015universal,lin2024power,harris2017quantum}. Each of these methods has its own challenges to overcome in order to achieve widespread adoption~\cite{fu2024optical,gupta2025neuromorphic}. However, a common requirement for all photonic computing approaches is that computing errors must remain low enough to avoid significant performance degradation~\cite{fang2019design,vadlamani2023transferable}, an issue that is central to analog computing.

Among various photonic computing approaches, integrated meshes of MZIs has received particular attention due to its scalable nature~\cite{shen2017deep} and universal programmability~\cite{hamerly2022accurate,hamerly2022stability,miller2013self,reck1994experimental,clements2016optimal,pai2020parallel,miller2017setting}. Despite these strengths, MZI-based photonic processors suffer from non-idealities such as unbalanced losses~\cite{clements2016optimal,hamerly2022accurate}, Digital-to-analog conversion (DAC) inaccuracies~\cite{hamerly2022accurate}, 50/50 beam splitter (BS) deviations~\cite{hamerly2022accurate,hamerly2022stability,hamerly2022asymptotically,miller2015perfect,pai2019matrix,russell2017direct,fang2019design}, unwanted phase shifts~\cite{hamerly2022accurate}, and reflections. Errors arising from unbalanced losses can be reduced at the hardware level by employing a rectangular mesh structure~\cite{clements2016optimal} in contrast to the triangular mesh design~\cite{reck1994experimental}.

To reduce the errors arising from variations in the 50/50 BSs, many other mesh architectures have been proposed~\cite{hamerly2022asymptotically,mojaver2023addressing,basani2023self,fldzhyan2020optimal,pai2019matrix,marchesin2025braided,fldzhyan2024low,miller2015perfect}, often at the cost of higher insertion losses. In addition, algorithmic approaches such as self-configuration~\cite{hamerly2022stability,hamerly2022accurate}, hardware error correction~\cite{bandyopadhyay2021hardware}, transferable learning~\cite{vadlamani2023transferable}, and in-situ backpropagation~\cite{hughes2018training,pai2023experimentally} have likewise been proposed to alleviate the 50/50 BS errors. These methods are tailored to specific errors and lack the flexibility to extend to other types of errors. Furthermore, no error-correction method has yet been proposed for errors stemming from unbalanced losses, DACs, and unwanted reflections, which can significantly degrade the performance of PMVMPs.

\begin{figure*}[htbp]
    \centering
    \includegraphics[width=0.92\linewidth]{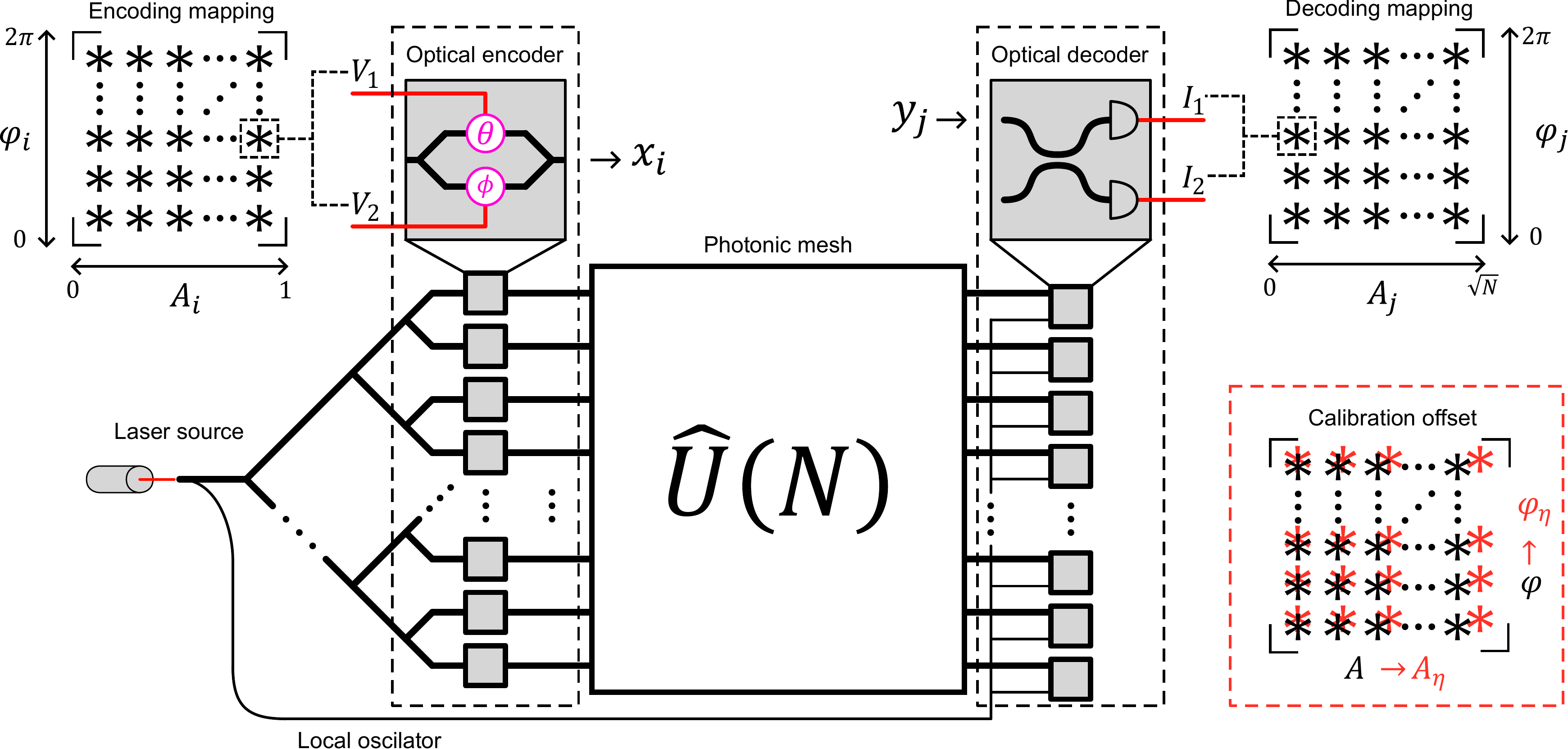}
    \caption{Schematic representation of a photonic matrix-vector multiplication processor consisting of an optical splitting tree that distributes the input laser source to an optical encoder, a photonic mesh, and an optical decoder. Vector data is encoded from digital information onto optical fields defined by the encoder mapping. The output optical vector is converted back into digital data defined by the decoder mapping. The en-/decoder mappings are shown here as black "$\ast$" representing voltage/current pairs as discrete points for simplicity. Scaling these mappings along the amplitude axis and shifting them along the phase axis corresponds to applying a complex factor $\eta$ to each in/output, providing 2N complex degrees of freedom for error correction.}
    \label{fig:PMVMP}
\end{figure*}

To conduct MVM in the optical domain, the vector data, $x_i$, stored in the electrical domain,  must be encoded into the amplitudes, $A_i$, and phases, $\varphi_i$, of the ingoing optical fields. MVM occurs as optical fields propagate and interfere throughout the photonic mesh, eventually reaching the output. The output optical fields, $y_j$, described by their amplitudes, $A_j$, and phases, $\varphi_j$, represent the MVM on the input vector by the photonic mesh and must be detected and converted back to the electrical domain for digital storage. This is done through decoders that can detect the optical signals and recover their amplitude and phase information. An illustration of such a PMVMP is shown in Fig.\ \ref{fig:PMVMP}. In this work, we introduce an error-correction method using optical encoders and decoders that are intrinsic to PMVMPs and demonstrate how it diminishes computational errors across various error sources and for different photonic mesh structures.
\section{Encoding mappings}
\noindent
Encoding can be achieved with various devices, such as an optical setup machine~\cite{miller2017setting,zhang2021optical,pai2023experimentally,zheng2024photonic}, off-chip modulators~\cite{xu2024large}, amplitude-only modulators~\cite{feldmann2021parallel,dong2024partial,moralis2024perfect}, directional-coupler-based complex encoders~\cite{qiu2024oplixnet}, a series of tunable couplers with phase shifters~\cite{bandyopadhyay2024single}, or as illustrated in Fig.\ \ref{fig:PMVMP}, a series of Mach-Zehnder modulators (MZM), which is widely employed in high-speed photonic computing ~\cite{xie2025complex,ahmed2025universal}. Regardless of the encoding method, correct encoding requires device calibration and, consequently, a function that maps the desired optical amplitudes and phases to corresponding encoder voltages.

\begin{figure*}[htbp] 
    \centering
    \includegraphics[width=0.65\linewidth]{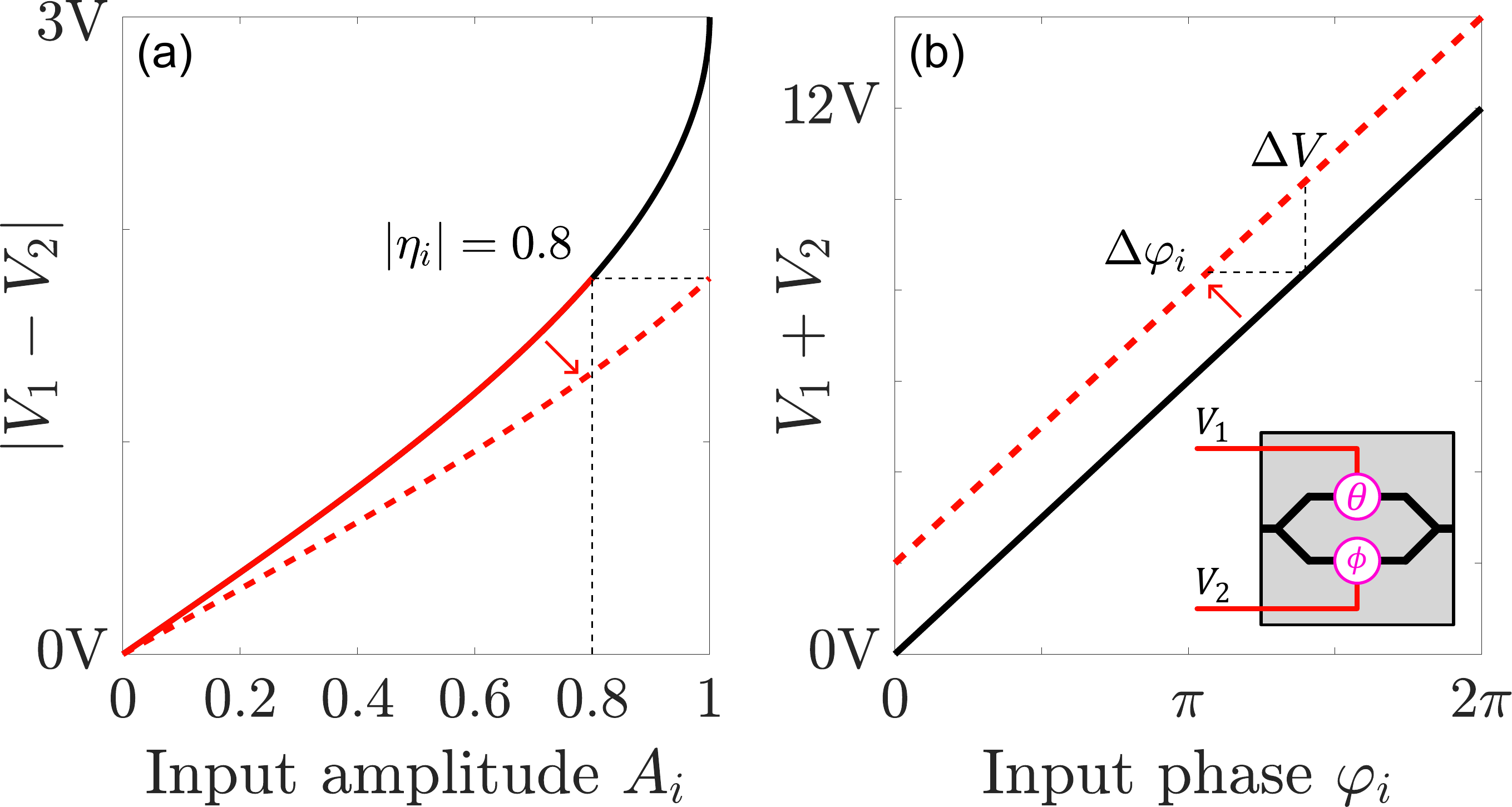}
    \caption{Example of a Mach-Zehnder modulator for encoding complex input data ($x_i=A_ie^{i\varphi_i}$) and its conversion mappings to tuning voltages ($V_1$ and $V_2$). (a) The amplitude is encoded from the difference in tuning voltages mapped by the solid line. Rescaling this mapping to the dashed line implements a correction factor of $|\eta_i|$ in amplitude (a correction factor of $|\eta_i|=0.8$ is chosen for illustrative purposes). (b) The phase is encoded from the sum of the tuning voltages mapped by the solid line. Shifting this mapping to the dashed line corresponds to adding a constant phase shift of, $\Delta\varphi_i$, to the encoded phase, $\varphi_i$. }
    \label{fig:Fig1.5}
\end{figure*}

A typical mapping between tuning voltages and encoded amplitude and phase of such a MZM device is illustrated in Fig.\ \ref{fig:Fig1.5}. The MZM consists of two Y-splitters and two phase shifters positioned in the interferometer arms. By tuning the relative phase between the arms, constructive or destructive interference is achieved, resulting in amplitude modulation of the output. 
Analog MZMs are typically fabricated in LiNbO3, owing to its strong Pockels effect and high-speed response ~\cite{wang2018integrated,xu2020high}. For such MZMs with a $\pi$-phase offset in one arm, the modulated amplitude can be represented by a sine function of the differential drive voltage $ A_i=\sin \left|\pi\frac{V_1-V_2}{2V_\pi}\right|$ (the solid curve in Fig.\ \ref{fig:Fig1.5}(a)). While this mapping reflects the underlying mechanism of MZM, it can be adjusted to accommodate the errors of the following photonic mesh. In doing so, we can choose a mapping such that the amplitude is scaled by a factor $|\eta_i|$, i.e., $A_i \to |\eta_i|A_i$ (the dashed line in Fig.\ \ref{fig:Fig1.5}(a)), resulting in a mapping from amplitude to voltage expressed by $|V_1-V_2|=\frac{2V_\pi}{\pi} \sin^{-1}|\eta_iA_i |$. 
This corresponds to a rescaling of the x-axis. Thus, the solid line in Fig.\ \ref{fig:Fig1.5}(a), which one would typically measure from device calibration, can be rescaled, forming the dashed line. We show later how this approach proves effective in reducing MVM errors. 

In addition to amplitude, the MZM also enables the encoding of the phase expressed as $\varphi_i=\pi\frac{V_1+V_2}{2V_\pi}$ (the solid line in Fig.\ \ref{fig:Fig1.5}(b)). Similar to amplitude, we can adjust the phase shift by translating the solid line left/right on the x-axis ($\Delta\varphi_i$), which can be implemented by simply adding a voltage bias $\Delta V/2$ to each arm. 
Thus, combined with the offset of the amplitude calibration, a complex correction factor, $\eta_i$, can be chosen to encode complex input data ($x_i\to\eta_ix_i$), providing extra degrees of freedom to mitigate photonic computing errors. This speaks to a general idea that any encoder can be represented by some mapping of voltage pairs to amplitude and phase, as illustrated by the black "$\ast$" in the top-left of Fig.\ \ref{fig:PMVMP}. Thus, by scaling this mapping along the amplitude axis and shifting it along the phase axis, we can implement our correction factor, $\eta_i$, as illustrated by the red "$\ast$" in the bottom-right of Fig.\ \ref{fig:PMVMP}.

\section{Decoding mappings}
\noindent
Decoding can be achieved with various schemes such as a reverse optical setup machine~\cite{miller2017setting, pai2023experimentally,zhang2021optical,zheng2024photonic}, an extra unitary layer~\cite{qiu2024oplixnet}, a 4-detector homodyne detection~\cite{xie2025complex}, or as illustrated in Fig.\ \ref{fig:PMVMP}, a decoder consisting of multiple homodyne detection (HD) channels, allowing measurement of field quadratures through interfering the output signal with a local oscillator~\cite{bandyopadhyay2024single}. Correctly decoding the output signal requires calibration of the decoder components and, consequently, a function that maps the measured photocurrents to the complex vector data, $y_j$.

\begin{figure*}[htbp] 
    \centering
    \includegraphics[width=0.85\linewidth]{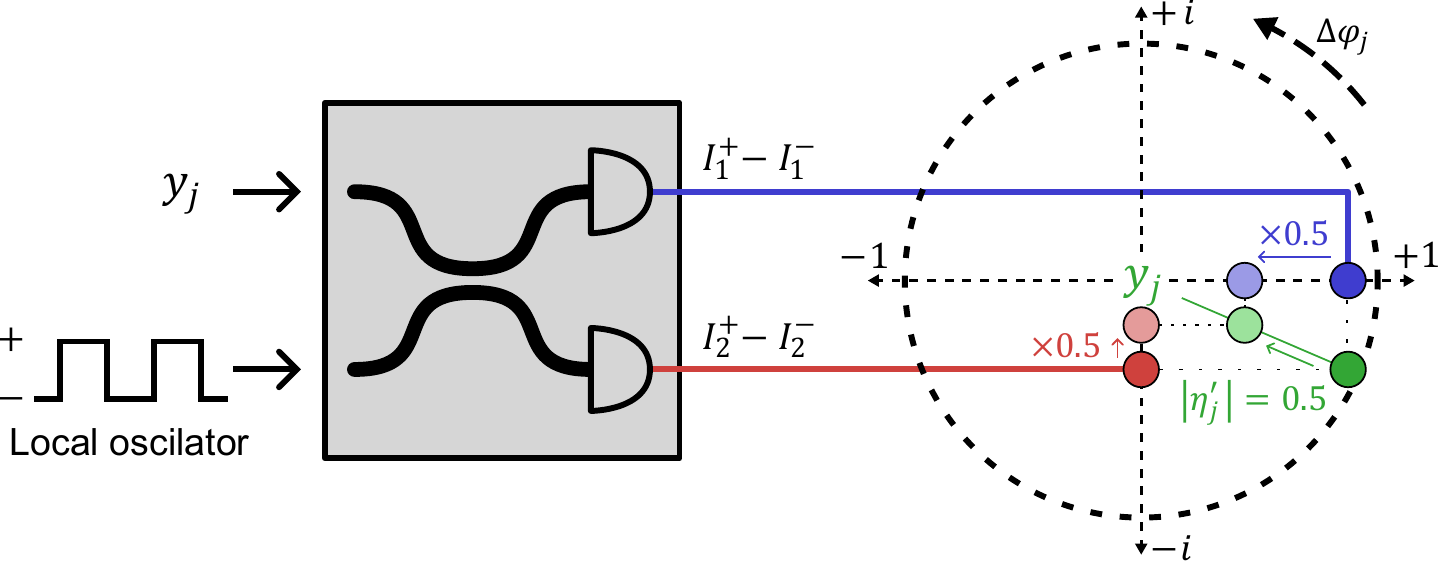}
    \caption{Example of a homodyne detection scheme for decoding, where the output data, $y_j$, is extracted by interfering with an oscillating reference signal (local oscillator), thus measuring the quadratures of the signal from the differences in photocurrents. Scaling of both the real and imaginary axes by some factor corresponds to implementing an amplitude correction factor exemplified here as $|\eta_j'|=0.5$. Implementing phase corrections can also be achieved by rotation of the complex plane.}
    \label{fig:Offset_decod}
\end{figure*}

A HD mapping between the measured photocurrents and the output vector data $y_j$ is illustrated in Fig.\ \ref{fig:Offset_decod}. After interference, the two fields are detected, generating subsequent photocurrents. If the local oscillator is phase-modulated (switching between $0$ and $\pi$), the real and imaginary parts of the data, $y_j$, can be measured, which are proportional to photocurrents ($\mathrm{Re}\{y_j\}\propto I_1^+-I_1^-$ and $\mathrm{Im}\{y_j\}\propto I_2^+-I_2^-$).
The proportionality constant then accounts for various factors such as signal strength of the local oscillator, optical losses, detector responsivity, uneven beamsplitting ratio, relative phase difference of the fanned-out local-oscillator branches, and phase variations due to fabrication imperfections, leading to a mapping from photocurrents to complex vector data. However, in order to accommodate the error of the preceding photonic network, one could choose other proportionality constants, $|\eta_j'|e^{\Delta \phi_j}$, such that the amplitude is scaled by a factor $|\eta_j'|$, i.e., $A_j\to|\eta_j'|A_j$, and the phase undergoes a phase change of $\Delta\varphi_j$, i.e., $\varphi_j\to\varphi_j+\Delta\varphi_j$. As illustrated in Fig.\ \ref{fig:Offset_decod}, this corresponds to rescaling the real and imaginary axes by $|\eta_j'|$ and rotating the complex plane by $\Delta\varphi_j$. This decoding mapping is ideal when the last-layer phase shifters of the PMVMP is implemented virtually through postprocessing~\cite{xie2025complex,shen2017deep,du2024ultracompact}. However, if PMVMP includes a physical last layer of phase shifters~\cite{zhang2021optical,zheng2024photonic,zhu2025versatile}, it may be preferable to apply phase correction ($\Delta\varphi_j$) directly through the last layer phase shifters and reserve decoding mapping only for amplitude correction. In both cases, we can implement a complex correction factor to each output data ($ y_j\to\eta_j'y_j$), providing additional degrees of freedom to mitigate errors. This approach is generalizable to any decoding device as illustrated in the top-right of Fig.\ \ref{fig:PMVMP}, where each "$\ast$" represents a pair (or more) of measured photocurrents that can be mapped to an amplitude and phase. Similar to the encoding mapping, by scaling this mapping along the amplitude axis and shifting it along the phase axis, we can implement our correction factor, $\eta_j'$, as illustrated in the bottom-right of Fig.\ \ref{fig:PMVMP}.

\section{Error-correction method}
\noindent 
As exemplified for a MZM in the former section, complex correction factors can be applied to a PMVMP through encoders, which effectively corresponds to applying a diagonal matrix, $D$, with correction factors, $\eta_i$, in front of the PMVMPs transformation, i.e., $\hat{U}\to\hat{U}D$. This likewise applies at the decoder level, equivalent to applying a diagonal matrix of correction factors, $\eta_j'$, in the back of the PMVMPs transformation, i.e., $\hat{U}\to D'\hat{U}$.

Equipped with these extra degrees of freedom, we seek to choose these correction factors such that we minimize the error in our MVMs.
The non-ideal matrix transformation $\hat{U}$ performed by the PMVMP can directly be measured by probing the system with canonical basis vectors, i.e., inputting vectors $[1,0,0,...,0], [0,1,0,...,0],..., [0,0,...,0,1]$ into the system. Thus, we attain $\hat{U}$, which can be compared against the ideal matrix transformation $U$. Due to non-idealities in the PMVMP, there will be a mismatch between these transformations, which we should seek to minimize. In Appendix \ref{app:Minerror}, we derive that to globally minimize the matrix error (defined from the Frobenius norm), the correction factors for the encoders should be found as:
\begin{equation}
\label{eq:eta}
    \eta_i=\frac{ \hat{v}^\ast_i \cdot v_i }{||\hat{v}_i||^2},
\end{equation}
where $\hat{v}_i$ and $v_i$ are the column vectors of $\hat{U}$ and $U$ respectively. Similarly, in order to find the optimal correction factors for the decoders, we can choose $\eta_j'$ according to Eq.~(\ref{eq:eta}), where $\hat{v}_j$ and $v_j$ are row vectors of $\hat{U}$ and $U$, respectively.

The encoder and decoder calibrations can be applied in tandem by simply applying the input correction, resulting in an updated matrix $\hat{U}\to\hat{U}D$, and then applying the output correction, or vice versa, for even stronger corrections. Implementing both corrections in tandem effectively yields $\hat{U}\to D'\hat{U}D$.

\begin{figure*}[htbp]
    \centering
    \includegraphics[width=0.9\linewidth]{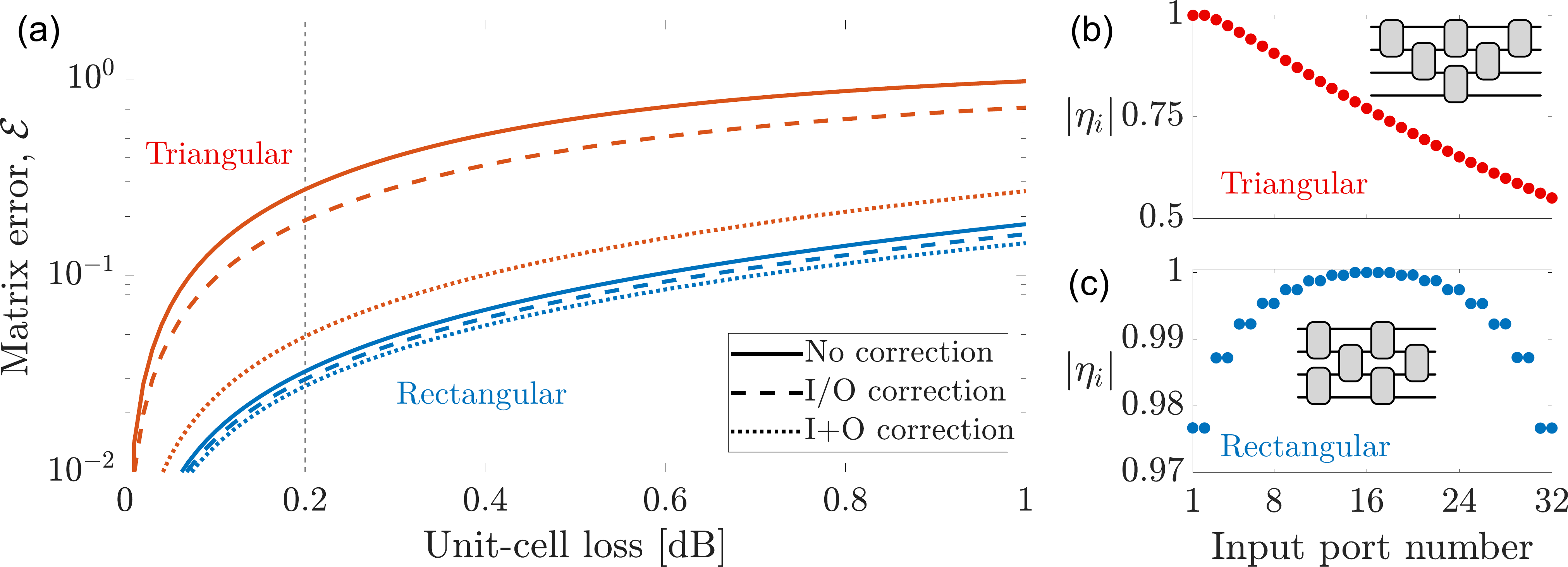}
    \caption{(a) The average matrix error for a $N=32$ universal multiport interferometer for the triangular mesh (red) or rectangular mesh (blue) with no error corrections (solid line), with error correction on either the encoder or decoder (dashed line) and with error correction on both the encoder and decoder (dotted line). (b-c) The normalized average correction factor amplitude for each input port for the triangular and rectangular mesh with a 0.2 dB unit-cell loss using both input and output correction. A small $N=4$ triangular and rectangular mesh is shown in each figure for conceptual clarity. The matrix error and correction factor have been averaged across $10^5$ Haar-random unitary matrices as a function of a constant loss of each unit cell in the mesh. The unit-cell loss is the only non-ideality in this simulation, and the non-ideal matrix has been normalized to account for the overall loss $\hat{U}\to \hat{U}/||\hat{U}||$. Note that the symmetric nature of the meshes leads to effectively the same performance for correcting on either the encoder or the decoder.}
    \label{Fig:MError}
\end{figure*}

\section{Performance}
\noindent
Loss remains a major challenge in photonic integrated circuits~\cite{Rosiek2023}, as light waves experience attenuation due to material absorption, scattering due to sidewall roughness, fabrication imperfections, and mode coupling. In PMVMPs, beyond reducing the overall signal-to-noise ratio, loss introduces another problem that is equally, if not more, important: its contribution to computational error in the form of unbalanced losses~\cite{clements2016optimal}. These errors can, in principle, be corrected by introducing additional sources of loss at the cost of increasing the overall complexity, footprint, and loss of the PMVMP~\cite{mojaver2023addressing,marchesin2025braided}. However, the error-correction method described in the previous section can easily be used to alleviate the unbalanced-loss errors without introducing additional components.

In order to assess the performance of our method on unbalanced-loss errors, we consider triangular and rectangular meshes~\cite{reck1994experimental, clements2016optimal}, with each unit cell subject to losses ranging from 0 to 1 dB. We consider an ensemble of unitary transformations randomly configured following the Haar measure~\cite{pai2019matrix,russell2017direct} and find correction factors to minimize the matrix error according to Eq.~(\ref{eq:eta}) for each transformation. We use the normalized definition of the Frobenius norm $||U||
^2=\frac{1}{N}\sum_{ij} |u_{ij}|^2$. This gives us a matrix error definition equivalent to the definition used in other works~\cite{hamerly2022asymptotically,hamerly2022accurate}:
\begin{equation}
    \mathcal{E}=||\hat{U}-U||,
\end{equation}
where we normalize our matrices to account for any overall loss such that $\hat{U}\to \hat{U}/||\hat{U}||$. For PMVMPs to reach real-world applications such as AI-acceleration, matrix errors must remain low to avoid significant performance degradation~\cite{vadlamani2023transferable,fang2019design}. Figure \ref{Fig:MError}(a) shows the average matrix error for ($N=32$) triangular and rectangular meshes. As shown, the matrix errors increase with the unit-cell loss for both meshes, while the triangular mesh exhibits significantly higher errors than the rectangular mesh, which would render it ineffective for most, if not all, processing tasks. This sets hard-to-meet requirements for unit-cell losses for the triangular mesh, as most state-of-the-art MZIs have insertion losses of around 0.1 to 0.2 dB~\cite{bandyopadhyay2024single,chen2012compact,suzuki2019low,harris2017quantum,taballione2021universal,taballione202320,kim2023programmable,zheng2023electro}. This vulnerability to unbalanced-loss errors can be alleviated by implementing a rectangular mesh, at the cost of more complex configuration algorithms. Nevertheless, even for the rectangular mesh, unbalanced losses still provide a non-negligible error contribution. 

Our error-correction method strongly reduces the unbalanced-loss errors as seen in Fig.\ \ref{Fig:MError}(a), with the strongest error reduction occurring when combining corrections to both the encoder (input) and the decoder (output). For unit-cell losses of 0.2 dB, this brings the average matrix error from 0.28 to 0.05 for the triangular mesh and from 0.033 to 0.028 for the triangular mesh, corresponding to 82\% and 15\% error reduction, respectively. The underlying mechanism can be understood through Fig.\ \ref{Fig:MError}(b-c), which shows the average amplitude correction factors of the encoders. The architecture of the triangular mesh leads to high propagation losses when passing through the upper I/O ports (low port numbers) and lower propagation loss for lower I/O ports (high port numbers). Thus, our error-correction method introduces larger amplitude correction factors, $|\eta_i|$, (closer to one) for low port numbers and smaller correction factors for high port numbers, as seen in Fig.\ \ref{Fig:MError}(b). In contrast, in a rectangular mesh, there are fewer unit cells at the top and bottom edges of the mesh. This leads to lower propagation losses when passing through the upper and lower I/O ports (high and low port numbers), thus, as seen in Fig.\ \ref{Fig:MError}(c), the error correction applies smaller correction factors to the upper and lower modulators and larger correction factors to the middle modulators to account for unbalanced losses. It should be noted that a rectangular mesh has a lower overall matrix error because the unit cells are distributed more uniformly across the mesh, leading to correction factors closer to 1. We do not show the phase correction factors, as they are quite small and on average zero for unbalanced-loss errors.

\begin{figure*}[htbp]
    \centering
    \includegraphics[width=0.75\linewidth]{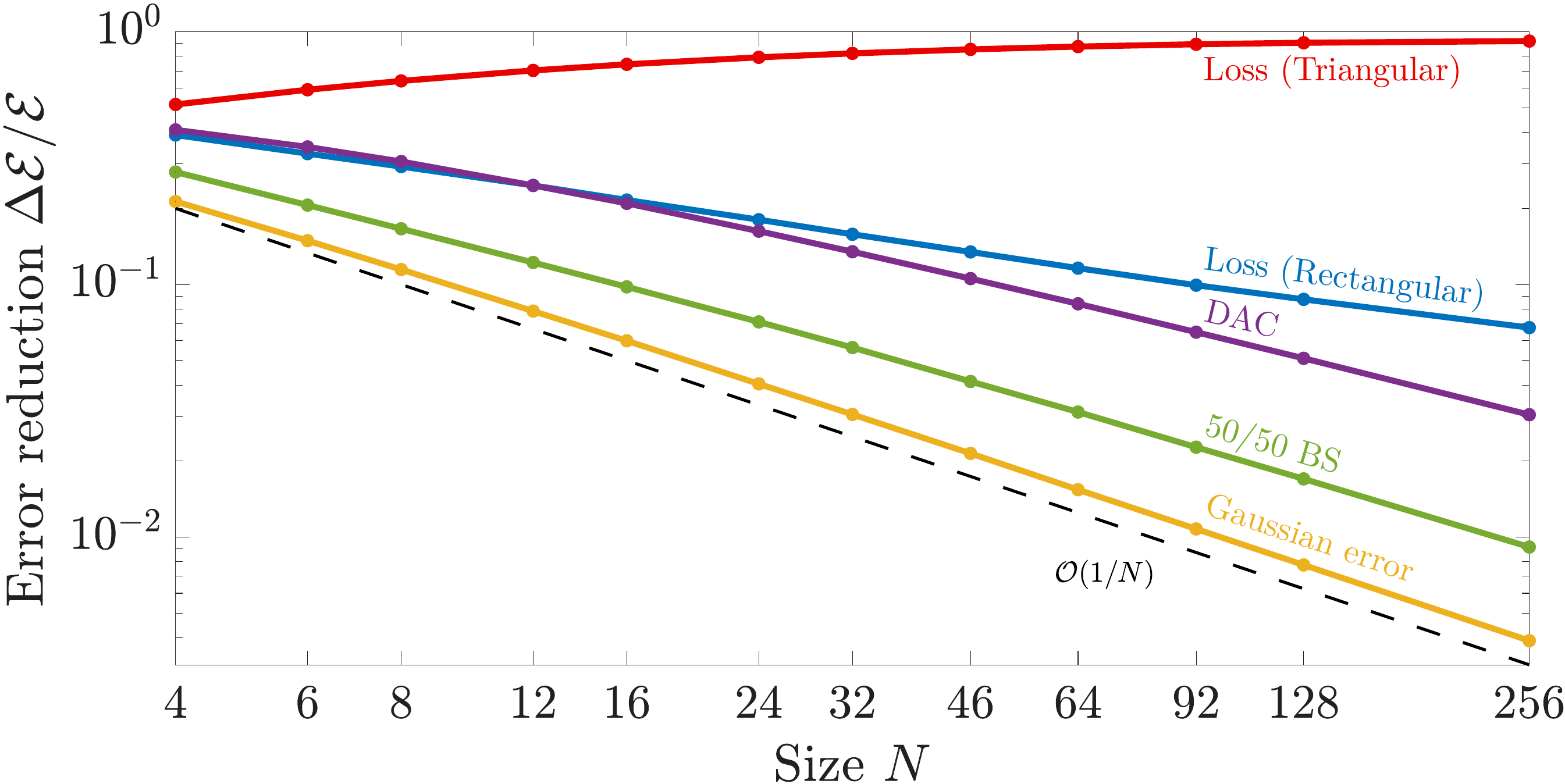}
    \caption{The average relative matrix error reduction for various non-idealities such as unbalanced losses in a triangular mesh (red), unbalanced losses in a rectangular mesh (blue), DAC errors (magenta), 50/50 beam splitter deviations (green), and Gaussian-distributed errors in matrix elements (yellow). To account for overall losses, the non-ideal matrices have been normalized $\hat{U}\to \hat{U}/||\hat{U}||$. The matrix errors are averaged over $10^3\times 256/N$ Haar-random unitary matrices. }
    \label{Fig:E_reduction}
\end{figure*}

To provide a broader picture of the application of our error-correction method beyond unbalanced-loss errors, and to investigate its performance as a function of the PMVMP size, we additionally consider other sources of error, such as 50/50 BS deviations and DAC inaccuracies, as well as Gaussian-distributed fluctuations in the matrix elements as a general model for MVM error. Figure \ref{Fig:E_reduction} shows the relative error reduction $\Delta \mathcal{E}/\mathcal{E}$ for various error types as a function of PMVMP size, where $\Delta \mathcal{E}$ is the difference between the corrected and uncorrected errors.

For the BS error, we consider a $2\%$ deviation from the ideal static 50/50 splitting ratio, consistent with values reported for foundries~\cite{mikkelsen2014dimensional}. Digital-to-analog conversion inaccuracies arise from the error introduced when a continuous signal between $0$ and $2\pi$ is quantized into $2^{B}$ discrete levels ($\sigma_{\mathrm{DAC}}=2\pi / \sqrt{12}\times 2^{-B}$)~\cite{hamerly2022accurate}. Similar to previous work, we assume DACs with 12-bit precision~\cite{hamerly2022accurate}. Both 50/50 BS and DAC errors have already been well studied~\cite{hamerly2022accurate,pai2019matrix,bandyopadhyay2021hardware,hamerly2022asymptotically}, and subsequent self-configuration algorithms have been developed to drastically reduce these errors~\cite{hamerly2022accurate,hamerly2022stability}. Despite this, our error-correction method yields a notable error reduction in both cases, with reductions of around 40\% for DAC and 30\% for 50/50 BS in small meshes to a few percent for large meshes, as shown in Fig.\ \ref{Fig:E_reduction}.

So far, we have described well-understood non-idealities as individual error sources. However, in PMVMPs, many other error sources whose explicit modeling is nontrivial may be present. Thus, in order to capture a general scenario, where multiple non-idealities and their interplay are present, we consider $1\%$ random Gaussian noise to the real and imaginary parts of each matrix element. For such random errors, we achieve approximately 20\% error reduction for small meshes and less than 1\% for large meshes.

Note that only unbalanced-loss errors depend on the mesh architecture, and other error sources are identical for both triangular and rectangular meshes. In addition, in most cases, the effectiveness of our method decreases with increasing PMVMP size, consistent with the trend observed in other error-correction methods~\cite{bandyopadhyay2021hardware,hamerly2022accurate,hamerly2022stability}. In our case, this reduction can be understood from the fact that unitary matrices have $N^2$ degrees of freedom (DOF), while our method only adds $2N$ new complex DOF, leading to a greater mismatch in DOF for larger PMVMPs. Thus, for uncorrelated matrix errors, we expect a scaling of $\mathcal{O}(1/N)$ in error reduction, which is exactly the scaling that we observe for Gaussian errors presented in Fig.\ \ref{Fig:E_reduction}. However, for other error types, we see a better scaling than $\mathcal{O}(1/N)$. We attribute this to correlated errors in the matrix elements, which can be corrected more effectively by the en-/decoder devices. This holds more strongly for unbalanced-loss errors, particularly for the triangular mesh, where we achieve around 50\% error reduction for small meshes and over 90\% for medium to large meshes. An interesting observation is that the relative error reduction for the triangular mesh increases with $N$. This distinctive characteristic arises from the triangular shape of the mesh, where the primary contributor to unbalanced losses is the variation in propagation loss when optical fields pass through different I/O ports. Since these differences in propagation loss can be readily corrected at the I/O ports where the encoders/decoders are present, our method delivers the most effective correction. Furthermore, as the network size increases, the unbalanced-loss error increases, resulting in a correspondingly larger relative error correction. Currently, the triangular mesh is only implemented for small PMVMPs with $N=4$ to $6$~\cite{carolan2015universal,zhang2021optical,pai2023experimentally,zheng2024photonic,zheng2023electro}. However, with the correction method presented in this work, large-scale PMVMPs can implement triangular meshes with errors that rival those of rectangular meshes.

\section{Discussion}
\noindent
Analog photonic computing relies on continuous wave quantities, i.e., amplitude and phase, thereby significantly affected by imperfections. As a result, error correction is essential for their performance. In this work, we have demonstrated that by offsetting the calibration of the encoder and decoder devices of a PMVMP, extra complex degrees of freedom can be attained to correct computing errors. In contrast to the rich body of existing work on error correction, our work presents a method that is applicable to a variety of errors, including unbalanced loss, DAC, 50/50 BS, and any unknown errors. For example, unwanted phase shifts and reflections at the interface between different components are common sources of errors in photonic integrated circuits. Although the precise modeling of these nonidealities is deferred to future work, the method we presented in this work can reduce errors arising from these and other unmodeled or unknown sources. The general-purpose nature of our error-correction method makes it appealing to a wide range of PMVMP architectures, such as unitary optical converters~\cite{tang2021ten,markowitz2024learning}, recirculating meshes~\cite{zhu2025versatile,perez2017silicon}, and even non-unitary architectures such as coherent cross-bar designs~\cite{moralis2024perfect,giamougiannis2023coherent} and singular-value decomposition meshes~\cite{xu2024large,miller2013self,tang2024lower}. In particular, photonic networks with predefined transformations that lack tunability~\cite{nikkhah2024inverse} may benefit from our work, and our method can prove pivotal in reducing fabrication-induced errors, as well as providing some level of programmability. Beyond its broad applicability, a key advantage of our work is that it is not restricted to unitary errors~\cite{hamerly2022stability} and proves very powerful against non-unitary errors, such as unbalanced losses, allowing for error reductions of more than 90\% for medium to large triangular meshes.

In this work, we have considered complex MVMs, as photonic networks naturally implement complex linear operations. However, there is a substantial body of work that traditionally performs real-valued MVMs using light, such as real-valued photonic tensor cores~\cite{ahmed2025universal,hua2025integrated,feldmann2021parallel,dong2024partial}, and diffractive optical processors~\cite{lin2018all,zhou2021large}. In such cases, our method still remains applicable by computing the correction factors in the same way using Eq.~(\ref{eq:eta}) but disregarding the phase-correction factors and only using the amplitude-correction factors, i.e., $|\eta_i|$ and $|\eta_j'|$. Furthermore, the extra degrees of freedom of encoder/decoder calibrations we introduced in this work could also be beneficial in all-optical neural networks~\cite{bandyopadhyay2024single}. These systems cannot differentiate between an ideal and a non-ideal transfer matrix operation. Thus, while optimal correction factors may not be very essential, the system can still be trained with these extra degrees of freedom.

\section*{Acknowledgments}
\noindent
We thank S. Stobbe for valuable discussions. We acknowledge financial support from the European Union’s Horizon research and innovation programme (Grant no. 101098961 – NEUROPIC) and the Innovation Fund Denmark (Grant No. 4356-00007B – EQUAL).

\appendix

\section{Minimization of matrix error}
\label{app:Minerror}

The Frobenius norm $||\cdot||$ of a matrix is canonically defined by the elementwise squared sum of the elements. This can also be written as the sum of the column vectors squared (we omit the $1/N$ factor for simplicity):

\begin{equation}
    ||A||^2=\sum_{i,j} |a_{ij}|^2=\sum_i ||w_i||_2^2.
\end{equation}
Thus, we can write the matrix error as the sum of vector differences :
\begin{equation}
    \mathcal{E}^2=||\hat{U}D-U||^2=\sum_i ||\hat{v}_i\eta_i-v_i||^2_2=\sum_i \mathcal{E}_i^2.
\end{equation}
In order to minimize the error, we simply have to minimize each of the vector errors $\mathcal{E}_i$. The 2-norm of a vector can be written as the inner product with itself $||w||^2=\langle w | w \rangle$. We can thus write our vector error as:
\begin{equation}
    \mathcal{E}_i^2=\langle v_i | v_i \rangle+|\eta_i|^2\langle \hat{v}_i | \hat{v}_i \rangle-\eta_i \langle v_i | \hat{v}_i \rangle-\eta_i^\ast \langle \hat{v}_i | v_i \rangle.
    \label{two-norm}
\end{equation}
Minimizing Eq.(\ref{two-norm}) can easily be done by first finding the phase such that $\arg (\eta_i)=\arg (\langle \hat{v}_i | v_i \rangle)$. Thereafter, the amplitude can easily be found to be:
\begin{equation}
    |\eta_i|=\frac{|\langle \hat{v}_i|v_i\rangle|}{||\hat{v}_i||^2}.
\end{equation}
By combining the amplitude and phase parts, we arrive at Eq.(\ref{eq:eta}). The same derivation can be done to minimize the decoder correction factors, $\eta_j'$, by instead considering the row vectors, $\hat{U}$ and $U$.

\bibliography{References.bib}

\end{document}